\documentclass[12pt,preprint]{aastex}
\usepackage{url}

\newcommand{\fig}[1]{Figure~\ref{#1}}

\newcommand{\speed}[1]{#1 km~s${}^{-1}$}

\newcommand{\acc}[1]{#1 m~s${}^{-2}$}
\newcommand{\rsun}[1]{${#1}\,R_\odot$}

\begin{document}

\shorttitle{} %

\shortauthors{Shen et al.}

\title{Evidence for the Wave Nature of an Extreme Ultraviolet Wave Observed by the Atmospheric Imaging Assembly Onboard the Solar Dynamics Observatory}

\author{Yuandeng Shen\altaffilmark{1,2}, and Yu Liu\altaffilmark{1}}

\altaffiltext{1}{Yunnan Astronomical Observatory, Chinese Academy of Sciences, Kunming 650011, China;\\
ydshen@ynao.ac.cn}
\altaffiltext{2}{Graduate University of Chinese Academy of Sciences, Beijing 100049, China}

\begin{abstract}
Extreme Ultraviolet (EUV) waves have been found for about 15 years. However, significant controversy remains over their physical natures and origins. In this paper, we report an EUV wave that was accompanied by an X1.9 flare and a partial halo coronal mass ejection. Using high temporal and spatial resolution observations taken by the {\em Solar Dynamics Observatory} and the {\em Solar-TErrestrial RElations Observatory}, we are able to investigate the detailed kinematics of the EUV wave. We find several arguments that support the fast-mode wave scenario: (1) The speed of the EUV wave (\speed{570}) is higher than the sound speed of quiet-Sun corona. (2) Significant deceleration of the EUV wave (\acc{-130}) is found during its propagation. (3) The EUV wave resulted in the oscillations of a loop and a filament along its propagation path, and a reflected wave from the polar coronal hole is also detected. (4) Refraction or reflection effect is observed when the EUV wave was passing through two coronal bright points. (5) The dimming region behind the wavefront stopped to expand when the wavefront started to become diffuse. (6) The profiles of the wavefront exhibited a dispersive nature, and the magnetosonic Mach number of the EUV wave derived from the highest intensity jump is about 1.4. In addition, triangulation indicates that the EUV wave propagated within a height range of about 60--100 Mm above the photosphere. We propose that the EUV wave observed should be a nonlinear fast-mode magnetosonic wave that propagated freely in the corona after it was driven by the CME expanding flanks during the initial period.
\end{abstract}

\keywords{Sun: corona -- Sun: oscillations -- Sun: flares -- Sun: magnetic topology -- Sun: coronal mass ejections (CMEs)}%

\section{INTRODUCTION}
The globally propagating wave-like disturbances in the solar corona were firstly reported by \cite{mose97}, and the detailed study was firstly presented by \cite{thom98,thom99}, based on the Extreme-ultraviolet Imaging Telescope \citep[EIT;][]{dela95} onboard the {\em SOlar and Heliospheric Observatory} \citep[{\em SOHO};][]{domi95}. Such coronal disturbances were dubbed EIT or EUV waves. The EUV waves generally appear as broad, diffuse features that form a single arc-shaped bright front on EUV difference images. They can propagate over most of the visible solar disk, and are often followed by expanding coronal dimming regions \citep{thom98}. The propagation speeds of the EUV waves are typically \speed{200--400} \citep{thom09}. Observational studies have demonstrated that some EUV waves tend to avoid magnetic structures (e.g., active regions, coronal holes, etc.) \citep[e.g.,][]{thom99,will99}. The EUV waves are usually observed to be generated at the periphery of active regions, and often associated with flares, coronal dimmings, Type II radio bursts, and coronal mass ejections (CMEs) \citep{bies02}. Statistical studies indicate that the EUV waves have a strong correlation with CMEs, but the relationship between them and flares as well as Type II radio bursts is relatively weak \citep{klas00,bies02,cliv05,chen06}. Despite extensive observational and theoretical research on the EUV waves, significant controversy remains over their physical mechanisms and origins.

Since the typical speeds of the EUV waves are higher than the quiet-Sun coronal sound speed ($c_{\rm s}\leqslant$ \speed{200}) and comparable to the local Alfv\'{e}n speed ($v_{\rm A}\leqslant$ \speed{1000}\citep{asch05}), and the EUV waves are often associated with flares and CMEs, therefore, they were initially interpreted as compressive fast-mode magnetosonic waves ($v_{\rm f}=\sqrt{c_{\rm s}^{2}+v_{\rm A}^{2}}$) driven by flares or CMEs \citep[e.g.,][]{thom98,thom99,thom00,will99,warm01,vrsn01,huds03,vrsn06,trip07}. Many authors propose that the EUV waves could be a good candidate of the counterpart of the H$\alpha$ Moreton waves \citep[e.g.,][]{more60} in the solar corona \citep{vrsn02,khan02,naru02,gilb04a,gilb04b,naru04,warm04a,warm04b,vrsn05,whit05,bala07}, which have been successfully explained as coronal fast-mode waves sweeping the chromosphere \citep{uchi68}. However, many observations demonstrate that the speeds of the EUV waves are often smaller than that of the Moreton waves. To reconcile this discrepancy, some authors argue that it can be explained by considering the deceleration of the EUV waves \citep{warm04a,warm04b}. In H$\alpha$ observations, the waves can only be observed during the early stage of the EUV wave. Whereas, in EUV observations, they can be traced to much longer distances. Therefore, the speeds measured in H$\alpha$ wavelengths are just the initial speeds of the EUV waves, which are still of high speeds. If the EUV waves propagate with a significant deceleration rate, the velocity discrepancy between the Moreton and EUV waves should be reasonable \citep{warm04a}. A number of observations indicate that the EUV waves do show the evidence for the deceleration of an order of several hundred m s$^{-2}$ \citep{vrsn02,vrsn05,warm01,warm04a,long08,vero08,pats09a,vero11}. Other characteristics in terms of a freely propagating fast-mode wave are also identified, such as the broadening of the perturbation profile and decreasing of the perturbation amplitude during the propagation period of the EUV waves \citep{warm04b,warm05b,vero10,long11a,long11b}. In addition, some observations show that the EUV waves are not always observed simultaneously in H$\alpha$ and EUV wavelengths, which might be resulted from the low magnetosonic Mach numbers of the EUV waves \citep{naru02,huds03,warm05b,warm11,kien11}. With wavelet analysis technique, \cite{ball05} showed an EUV wave with a well defined period of about 400 s and a minimum energy of $\sim 3.4 \times 10^{18}$ J, which is comparable to the energy of a nanoflare.
Occasionally, the EUV waves can occur repeatedly within a short timescale from the same source region \citep{kien11,zhen12}. For these homologous EUV waves, it is propitious to perform a quantitative analysis of the characteristic wave parameters without any limitation like changing or unknown quiet-Sun background conditions.

The EUV waves have been observed to interact with various coronal magnetic structures (e.g., filaments and coronal holes) along their propagation paths \citep{thom98,thom99,will99,eto02,vero06,gilb08,gopa09,hers11,asai12,li12}. \cite{thom99} and \cite{trip07} reported on the EUV waves that were stopped by remote coronal holes as well as near the separatrix between active regions, where they may appear as a ``stationary'' front \citep{dela99}. These observational characteristics have been confirmed by a few numerical simulations \citep{wang00,wu01,ofma02,schm10}. \cite{gopa09} found the evidence for an EUV wave that was reflected by a remote coronal hole, implying that the EUV transients are possibly a true wave. \cite{liu10} found an EUV wave that simultaneously exhibited both fast and slow wave components. In their case, the preceding slow component was overtaken by the following fast component one, and the interaction between them not only produced multiple ripples but also steeped the local amplitude of the wave. This behavior is the characteristic of a real wave. The EUV waves were also observed to interact with remote filaments, which usually caused the oscillations of filaments \citep{eto02,gilb08,asai12}. Stereoscopic observations taken by {\em Solar-TErrestrial RElations Observatory} \cite[{\em STEREO};][]{kais08} were used to study the three-dimensional kinematics of EUV waves \citep{long08,vero08,ma09,kien09,pats09a,temm11}, which improved the characterization of the EUV waves due to its relatively high temporal and spatial resolution and multi-angle observations. Triangulation studies indicate that the EUV waves are trapped in a certain layer of the solar atmosphere ranging from 80 to 100 Mm above the photosphere \citep{pats09b,kien09}. This height range is comparable to the coronal scale-height of 50--100 Mm for quiet-Sun temperatures of 1--2 MK. In addition, the evidence of angular rotation for the EUV waves during propagation has also been reported \citep{podl05,attr07}. When EUV waves occur on the disk limb, they often exhibit a dome-shaped structure that enables us to further study the relationship between the EUV waves and the associated CMEs. So far several such observations have been documented \citep{vero10,chen12,selw12}, and the authors have shown that the expanding CMEs are preceded by the EUV waves, and they thereby propose that the EUV waves are driven by the expanding CMEs. It is worth noting that the EUV waves have been applied on coronal seismology to estimate the global properties of the solar corona, assuming that the EUV waves are true waves in nature \citep[e.g.,][]{warm05a,west11,long11a}.

Although a lot of extensive observational and theoretical works have reached the conclusion that fast-mode wave models can account for the EUV waves \citep[e.g.,][]{thom98,pats09a,warm04a,warm04b,muhr10,vero10,long11a,long11b,zhen11,chen12}, many other studies show that the EUV waves are not at all waves in a physical sense \citep[e.g.,][]{dela99,dela00,fole03,harr03,attr09,attr10,zhuk09,dai10,chen11}. \cite{dela99,dela00} proposed that the EUV waves are more related to the changes in magnetic topologies due to the eruption of CMEs than to real waves. In their following works, they proposed that the EUV waves could be interpreted as the Joule heating of the coronal plasma at the interface between the CME legs and the surrounding quiet-Sun magnetic field \citep{dela07,dela08}. Stimulated by the findings of \cite{dela99} and \cite{dela00}, a numerical field-line stretching model was also proposed to explain the EUV waves \citep{chen02,chen05}, which is based on the erupting of a flux rope and predicts a very fast super-Alfv\'{e}nic piston-driven shock preceding a slower pseudo-wave that is associated with the successive opening of the overlying magnetic field induced by the erupting flux rope. In this model, the authors argued that the fast shock might be the coronal counterpart of the Moreton wave observed in H$\alpha$ wavelengths, while the slow pseudo-wave behind the shock corresponds to the EUV wave observed in the corona. Several alternative models have also been proposed to interpret the EUV waves, such as the successive reconnection model \citep{attr07,van08}, the slow-mode wave model \citep{will06,wang09}, and the scenario that the EUV waves could be interpreted as soliton waves \citep{will07}. In addition, some authors proposed that both wave and non-wave models are required together to explain the observations and understand the complex natures of the EUV waves \citep[e.g.,][]{zhuk04,cohe09,liu10,down11,li12}. Detailed information for the characteristics and interpretations of the EUV waves can be found in several recent papers \citep[see][]{will09,warm10,gall11}.

Up to the present, both the wave and non-wave theories could not fully interpret the EUV waves. To clarity the physical nature of the EUV waves, more detailed observations with high temporal and spatial resolution data as well as theoretical studies are desirable at the present. In this paper, we present an EUV wave that occurred on September 24, 2011. It was accompanied by a {\em GOES} soft X-ray (SXR) X1.9 flare and a fast partial halo CME from the active region NOAA 11302 (N13E46). With the high temporal and spatial resolution observations taken by the {\em Solar Dynamics Observatory} ({\em SDO}) and {\em STEREO}-B instruments, we are able to analyze the wave kinematics in a great detail, seeking to find more new clues to clarify the physical natures of the EUV waves. We find several clues supporting the scenario that the EUV waves are propagating fast-mode waves in the corona. Instruments and data sets are described in Section 2, results are presented in Section 3, discussion and conclusions are given in Section 4.

\section{INSTRUMENTS AND DATA SETS}
The EUV wave occurred near the east disk limb, which means that the on-disk EUV wave could be observed from both {\em SDO} and {\em STEREO}-B directions. The Extreme Ultraviolet Imager \citep[EUVI;][]{wuel04} onboard {\em STEREO}-B has observed this event in four channels: 171, 195, 284, and 304 \AA. We only use the 195 \AA\ observations in this paper, which recorded the EUV wave with only three frames. The cadence and pixel resolutions of 195 \AA\ channel are 5 minutes and $1\arcsec.6$, respectively. The Helioseismic and Magnetic Imager \citep[HMI;][]{scho12} onboard {\em SDO} provides line-of-sight magnetograms at a 45 s cadence with a precision of 10 Gauss. The Atmospheric Imaging Assembly \citep[AIA;][]{leme12} onboard {\em SDO} has high time cadences up to 12 s and high pixel resolution of $0\arcsec.6$. It captures images of the Sun's atmosphere out to \rsun{1.3} in seven EUV and three UV-visible wavelengths. We mainly focus on the AIA observations to investigate the kinematics of the EUV wave presented in this paper due to its higher temporal and spatial resolution. On September 24, 2011, the separation angle between {\em STEREO}-B and {\em SDO} was about 96${}^{\circ}$. In addition, the {\em Reuven Ramaty High Energy Solar Spectroscopic Imager} \citep[{\em RHESSI};][]{lin02} hard X-ray (HXR) count rates and the {\em Geostationary Operational Environmental Satellite} ({\em GOES}) SXR fluxes are also used. All images used in this paper are differentially rotated to a reference time (09:40:00 UT), and the solar north is up, west to the right.

\section{RESULTS}
\begin{figure}\epsscale{0.5}
\plotone{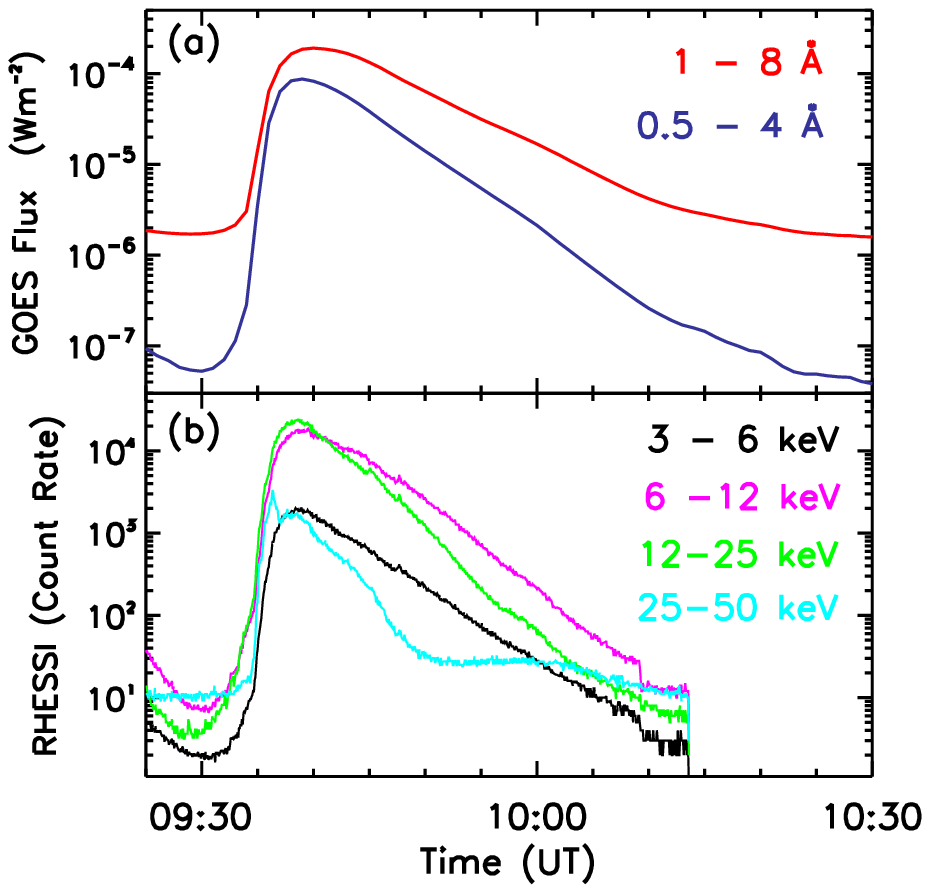}
\caption{Panel (a): Time profiles of {\em GOES} 0.5--4 \AA\ (blue) and 1--8 \AA\ (red) SXR. Panel (b): The {\em RHESSI} HXR count rates in the energy bands (4 seconds integration) of 3--6 keV (black), 6--12 kev (magenta), 12--25 keV (green), and 25--50 keV (cyan). \label{fig1}}
\end{figure}

The EUV wave was associated with a {\em GOES} X1.9 flare in NOAA AR11302 (N13E46) and a fast partial halo CME was observed by the Large Angle and Spectrometric Coronagraph Experiment \citep[LASCO;][]{brue95} onboard {\em SOHO}. The EUV wave was observed simultaneously by both {\em SDO} and {\em STEREO}-B from different viewpoints. The start, peak, and end times of the flare are 09:30, 09:39, and 10:09 UT, respectively. The fluxes of {\em GOES} SXR and {\em RHESSI} light curves in several energy bands are plotted in \fig{fig1} to show the flare. This flare was very energetic, and it appeared to be a white-light flare in the HMI continuum intensity observations. According to the SOHO LASCO CME CATALOG\footnote{http://cdaw.gsfc.nasa.gov/CME\_list}, the CME's center was at a position angle (PA) 92${}^\circ$, and its angular width was 145${}^\circ$. The application of a linear fit to the height-time points measured by S. Yashiro yields the speed of the CME to be about \speed{1936}, while the acceleration from quadratic fit is about \acc{-25}. These results confirm that fast CMEs are often associated with flares and coronal waves on the visible disk \citep{shee99}. In this paper, we mainly focus on the EUV wave using the EUV observations taken by {\em SDO}/AIA due to the high temporal and spatial resolution observations.

\begin{figure}\epsscale{0.6}
\plotone{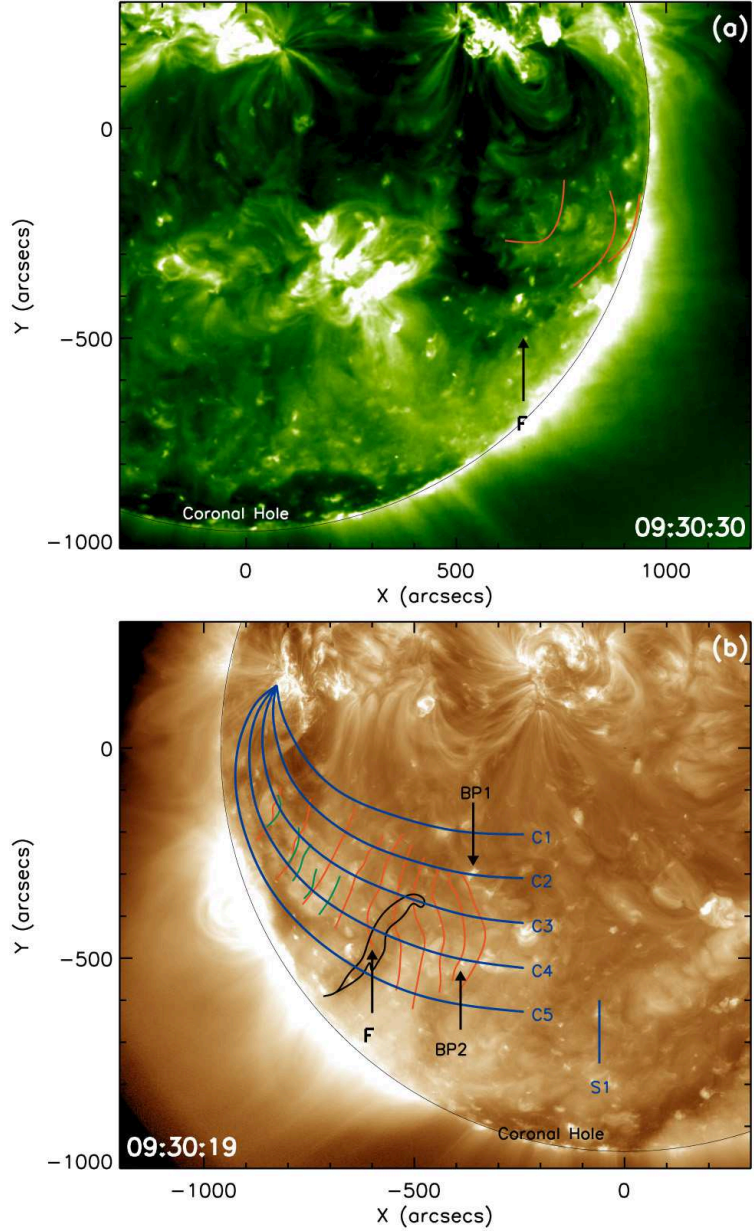}
\caption{{\em STEREO}-B EUVI 195 \AA\ (a) and AIA 193 \AA\ images (b) showing the coronal condition where the EUV wave propagates. The cyan (orange) curves mark the wavefronts of W1 (W2), which are detected from a series of base-difference images. The blue curves C1 and C5 mark the region where the EUV wave propagates, and we get time-distance diagrams from C2--C4. The blue line (S1) is used to detect the reflected wave from the polar coronal hole. The profile of the filament (black contour) is overlaid on the AIA 193 \AA\ image. BP1 and BP2 are two bright points on C2 and C4, respectively. The FOV is $1500\arcsec \times 1300\arcsec$. An animation of the EUV wave is available in the online version of the journal. \label{fig2}}
\end{figure}

An overview of the EUV wave event is shown in \fig{fig2}. The propagation of the EUV wave was restricted within a conical sector of the quiet-Sun corona, in which a filament and a few bright points can be identified (indicated by the black arrows in \fig{fig2} (a) and (b)). Before the appearance of the primary wave (W2), there was another EUV wave (W1) preceding W2, which can also be observed from the AIA 193 \AA\ and 211 \AA\ running difference images. The leading edges of the wavefront of the propagating EUV waves (determined from the AIA 193 \AA\ base-difference images) at different times are overlaid on the {\em STEREO}-B 195 \AA\ image (\fig{fig2}(a)) and the SDO AIA 193 \AA\ image (\fig{fig2}(b)), where the cyan (yellow) curves represent the leading edge of the wavefront of W1 (W2), respectively. In \fig{fig2}(b), the blue curves perpendicular to the wavefront are used to obtain the time-distance diagrams for analyzing the detailed kinematics of the EUV waves, while the vertical blue line (S1) is used to detect the reflected wave from the south polar coronal hole. Note that the blue curves used to obtain the time-distance diagrams are not the great circles passing through the flare site. This technique can minimize the spherical projection effect. However, since the EUV wave presented here did not propagate along the great circles passing through the flare kernel, we define the paths that are perpendicular to the wavefronts to obtain the time-distance diagrams, which should reflect the kinematics projection for the wave in the plane of the sky.

\begin{figure}\epsscale{0.8}
\plotone{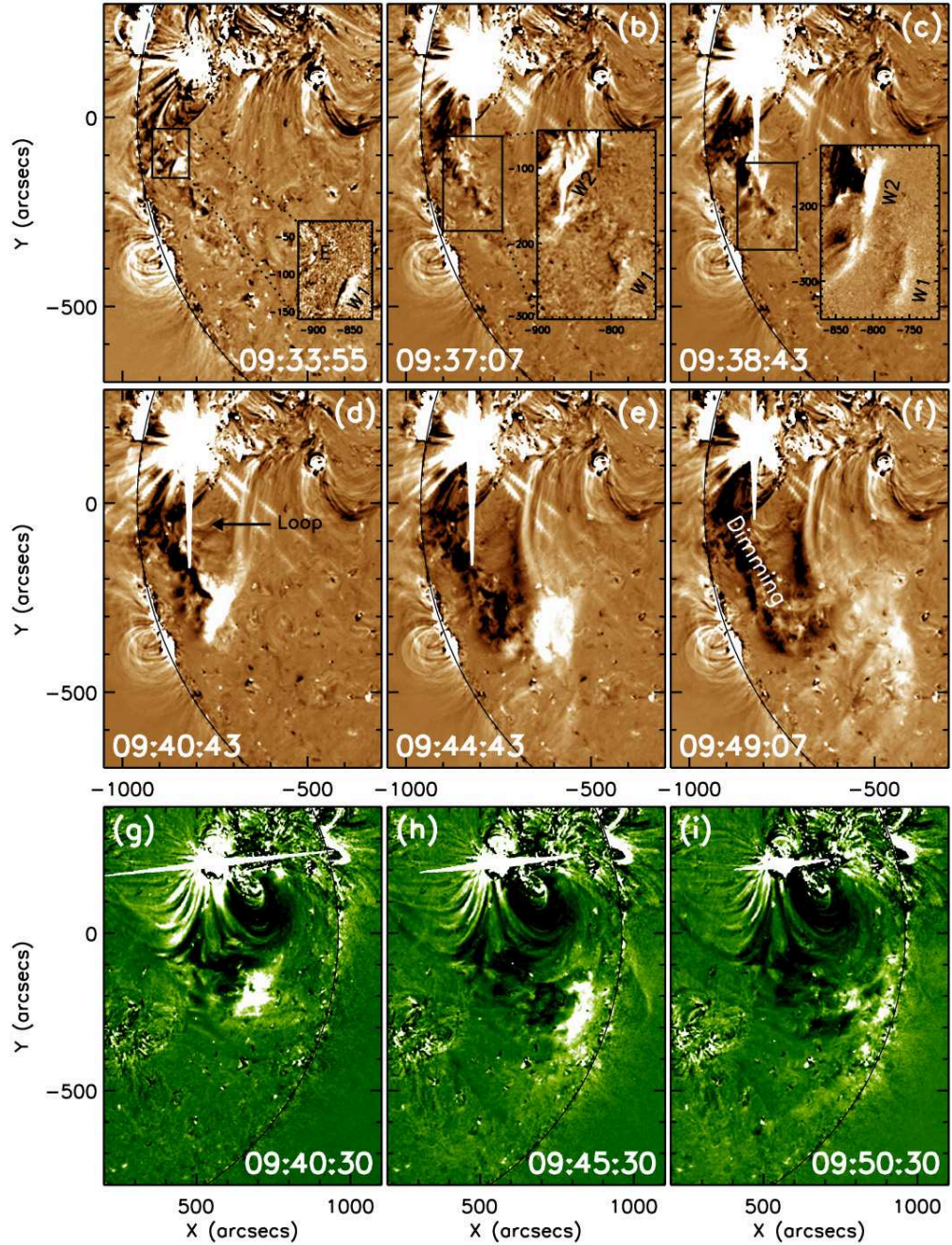}
\caption{AIA 193 \AA\ (a--f) and {\em STEREO}-B EUVI 195 \AA\ (g--h) base-difference images showing the EUV wave. The black rectangle regions in panels (a--c) are magnified as shown by the insets in the same panel. It should be noted that the images displayed in the insets are the AIA 193 \AA\ running difference images. The black arrow in panel (d) points to one loop that showed obvious oscillation motion after the passage of the EUV wave. \label{fig3}}
\end{figure}

We show the morphological evolution of the EUV wave in \fig{fig3}. Note that the insets are the AIA 193 \AA\ running difference images, which show the EUV waves clearly. Before the start of the X1.9 flare, a small plasma ejection can be identified on the AIA 193 \AA\ base-difference images. Several minutes later, a weak EUV wave (W1) preceding the plasma ejection can be observed at about 09:33:55 UT (see the inset in \fig{fig3}(a)). This wave was possibly driven by the small plasma ejection behind it. At about 09:37:07 UT, the primary EUV wave (W2) appeared clearly on the AIA 193 \AA\ difference images, which was preceded by W1 (see the inset in \fig{fig3}(b)). Obviously, the propagation of W2 was faster than that of W1, which can be identified by comparing the distance between them at different times (see the insets in \fig{fig3} and Animation1 available in the online version of the journal). During the initial stage of W2, an expanding dimming region behind the wavefront formed following the propagation of W2. The W2 wavefront during this period broadened gradually but still remained rather compact, while the propagation direction changed gradually from southeastward to westward. However, after or around 09:44 UT, while the dimming region stopped its expansion, the EUV wave still continued its propagation to an increasing distance up to the central meridian of the Sun. During this time period, the shape of the wavefront became more diffuse and the wavefront broadened a lot as seen in panel \fig{fig3}(f). These observational results suggest that the EUV wave should have been driven by the CME expanding flanks as proposed in several studies \citep{pats09a,muhr10,temm11}, in which the coronal dimming region should map the CME footprint on the solar surface, and the EUV wave can start to propagate freely after the driving of the CME expanding flanks. On the other hand, {\em STEREO}-B also catched the EUV wave, and we show the 195 \AA\ base-difference images in \fig{fig3}(g)--(i). By comparing the propagating wave in the AIA and {\em STEREO}-B observations, one can see that this wave was a surface wave that was probably trapped within a certain height in the corona. Triangulation of the EUV wave (using the procedure ``scc\_measure.pro'' developed by W. Thompson) indicates that the propagation height of this EUV wave ranges from 60 to 100 Mm above the photosphere, in agreement with the results reported by \cite{pats09b} and \cite{kien09}.

\begin{figure}\epsscale{0.8}
\plotone{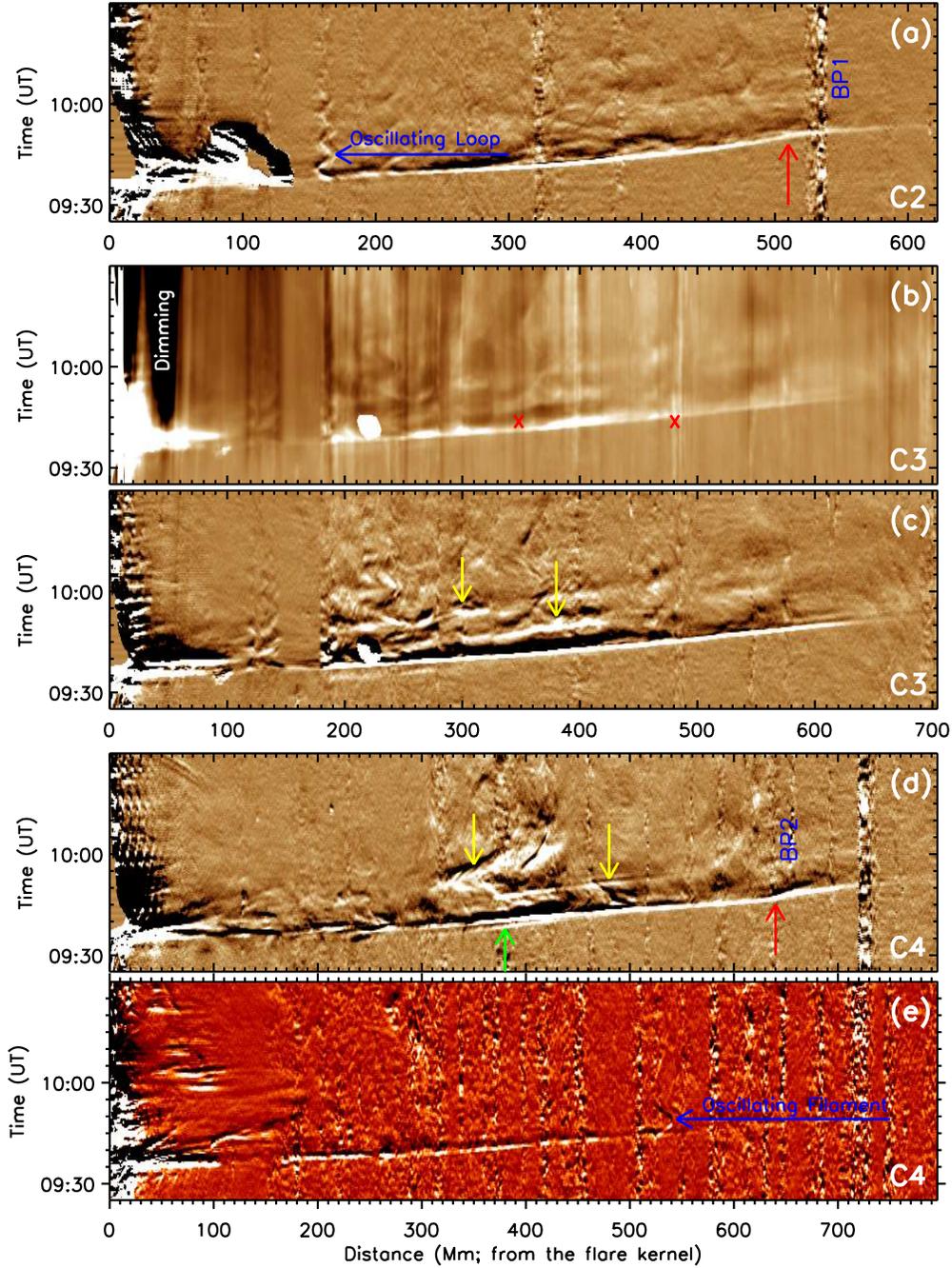}
\caption{Time-distance diagrams showing the kinematics of the EUV wave. Panel (a), (c), and (d) are obtained from AIA 193 \AA\ running difference images along the curves C2, C3, and C4, respectively. Panel (b) is obtained from AIA 193 \AA\ base-difference images along C3, while panel (e) is made from AIA 304 \AA\ running difference images. The blue arrows in panel (a) and (d) point to the oscillating loop and filament, respectively. The green arrow in panel (d) indicates the preceding W1. The red arrows point to the sites where the EUV wave speed changed significantly when passing through the bright points. Also, the yellow arrows indicate some secondary waves after the passage of the EUV wave (W2). \label{fig4}}
\end{figure}

\begin{figure}\epsscale{0.8}
\plotone{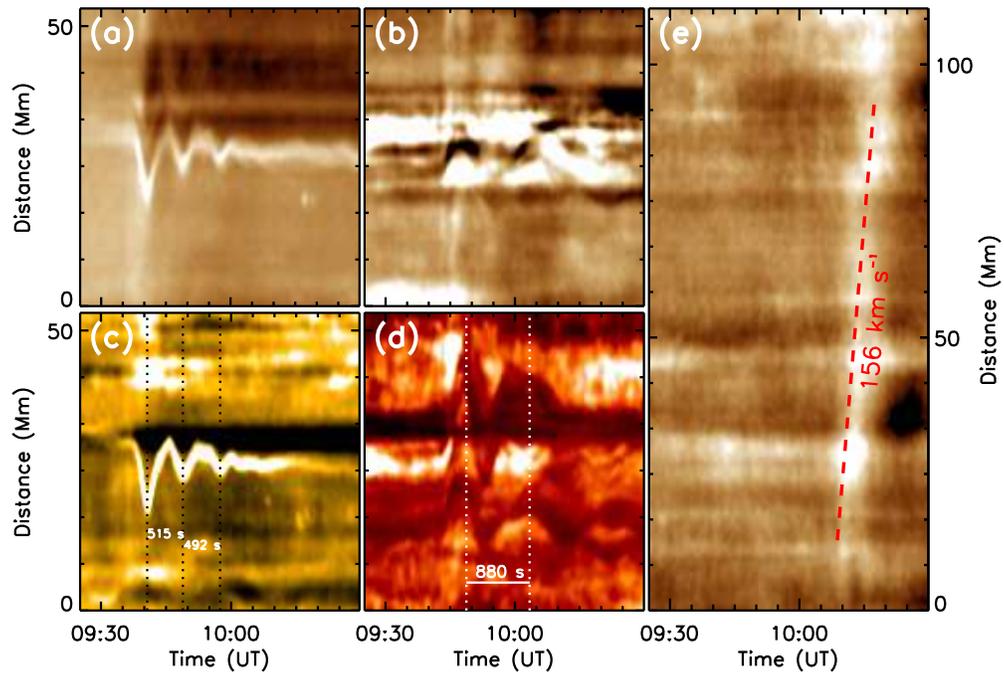}
\caption{AIA 193 \AA\ ((a), (b)), 171 \AA\ (c), and 304 \AA\ (d) base-difference images showing the oscillating loop ((a) and (c)) and filament ((b) and (d)). Panel (e) is the time-distance diagram along the blue line S1 as shown in \fig{fig2}(b). \label{fig5}}
\end{figure}

We study the kinematics of the EUV wave using the time-distance diagrams obtained along curves C2--C4 as shown in \fig{fig2}(b). The EUV wave could be observed in all EUV channels of AIA across a wide temperature range and it showed similar kinematics in these EUV channels. Here, we only show the EUV wave using the time-distance diagrams obtained from AIA 193 and 304 \AA\ base- and running difference images (see \fig{fig4}). In each time-distance diagram, the EUV wave displays as a slightly curved bright ridge. The slope of the bright ridge, which is defined as the ratio of the distance to the time within a limited period, represents the propagation speed of the EUV wave. One can see that the EUV wave could only be observed at a minimum distance of about 170 Mm from the flare kernel. In the time-distance diagram along C2, an oscillating loop can be identified after the passing of the EUV wave (see the blue arrow in \fig{fig4}(a) and the black arrow in \fig{fig3}(d)). This oscillating loop can be best seen in the AIA 171 \AA\ base-difference time-distance diagram (see \fig{fig5}(c)). It oscillated for several cycles before totally damping, and the oscillation period is about 500 s. When the EUV wave was passing through BP1, it interacted with BP1 and its propagation speed changed significantly (see the red arrows in \fig{fig4}(a)). The same effect can also be identified in the time-distance diagram along C4, where the speed of the EUV wave also changed obviously after its interaction with another bright points (BP2) (see the red arrows in \fig{fig4}(d)). The changes of the EUV wave speed might be caused by the reflection or refraction of the EUV wave at the bright points, indicating the wave nature for this EUV wave.

Both time-distance diagrams obtained from AIA 193 \AA\ base- and running difference images along C3 are shown in \fig{fig4} (b) and (c), respectively. From the base-difference time-distance diagram, one can see that the dimming region stopped its expansion at around 09:44 UT and lasted for a long time. The outer edge of the dimming region kept at a distance of about 50 Mm from the flare kernel. In the meantime, the EUV wave propagated out continuously before indiscernible on the visible disk. Besides the obvious primary wave stripe observed in the running difference time-distance diagrams, there are a few secondary wave tracks that could also be identified in the time-distance diagrams obtained from C3 and C4 (see the yellow arrows in \fig{fig4}). By checking the time-lapse movies made from the AIA observations, we find that these secondary waves were probably excited by some coronal structures when they interacted with the primary EUV wave, unlike the slow pseudo-wave predicted in the numerical model by \cite{chen02}. Along C4, the wave stripe that represents W1 can be identified in the AIA 193 \AA\ running difference time-distance diagram (see the green arrow in \fig{fig4}(d)). The speed of W1 was obviously slower than that of W2, and it was overtaken by W2 at about 09:40 UT. After that W1 was undetectable anymore. We show the interaction of the EUV wave with the filament using the AIA 304 \AA\ running difference time-distance diagram (\fig{fig4}(e)). After the passing of the primary EUV wave (W2), the filament started to oscillate for a few cycles but never erupt (see the blue arrows in \fig{fig4}(e) and \fig{fig5}(d)). The period of the oscillating filament is about 880 s. It is worth noting that the reflection of W2 by the south polar coronal hole was also observed, and we show the reflected wave in \fig{fig5}(e), which propagated northward at a speed of about \speed{156} in the plane of the sky. Here, we failed to detect the incident W2 and the interaction of W2 with the coronal hole in the time-distance diagram as shown in \fig{fig5}(e), since W2 had become very diffuse when it reached the coronal hole. However, the start time of the possibly reflected wave was consistent with the estimated arrival time of W2 to the coronal hole, and the time-lapse movie made from the AIA 193 \AA\ observations confirms the reflected wave from the polar coronal hole. Thus we believe that the detected wave along path S1 was the reflected wave of W2 from the polar coronal hole.

\begin{figure}\epsscale{0.5}
\plotone{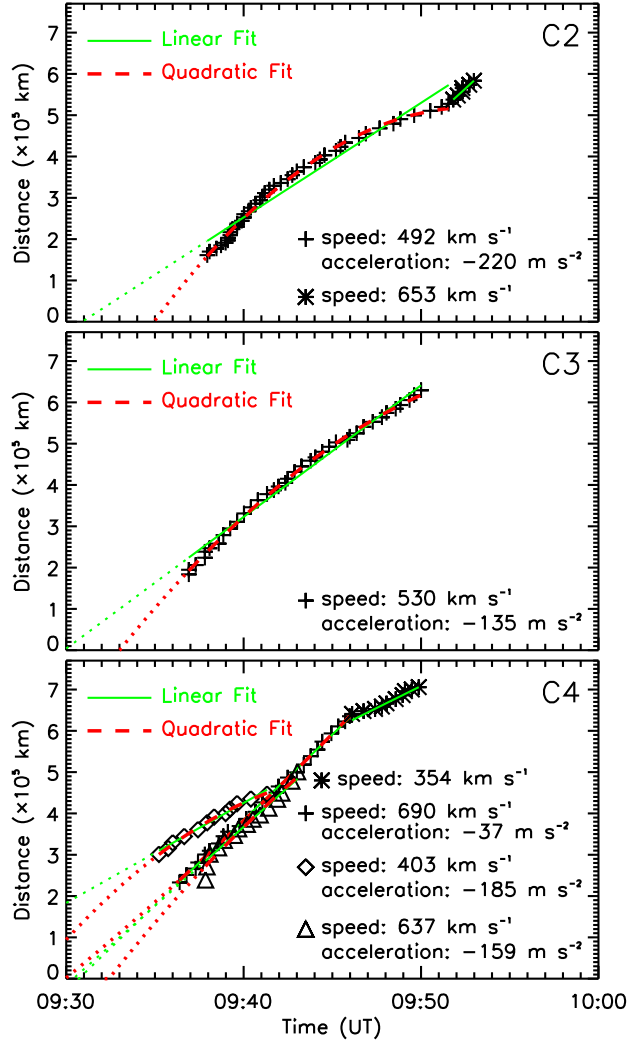}
\caption{Kinematics of the EUV wave. In each panel, the green (red) lines are the linear (quadratic least-squares) fit to the data sets that determined by visually tracking the white stripes from the time-distance diagrams as shown in \fig{fig4}, while the dotted section of each curve is a back-extrapolation of the fitted lines. Symbols ``+'' and ``$\ast$'' represent the main part and reflected part of the EUV wave, respectively. The symbol ``$\bigtriangleup$'' in the bottom panel represents the wavefront of the EUV wave in the 304 \AA\ time-distance along C4, and the symbol ``$\diamond$'' represent the preceding wave (W1) along C4. The speeds and accelerations of the EUV waves along each cut are also given in the figure. \label{fig6}}
\end{figure}

The speeds and accelerations of W1 and W2 in the plane of the sky are plotted in \fig{fig6}, in which the values of distance and time are determined from the time-distance diagrams shown in \fig{fig4}. The speeds and accelerations are obtained by making a linear (green line) and quadratic (red line) fit to the data sets respectively. The dotted part of each fitted line is a back-extrapolation of the fitted lines. As seen in this figure, the propagations of both W1 and W2 exhibit a significant deceleration along each cut. The speed (acceleration) of W1 is about \speed{403} (\acc{-185}). For W2, its speed ranges from \speed{492} to \speed{690} and the average value is about \speed{570}. The acceleration ranges from \acc{-37} to \acc{-220} and the average value is about \acc{-130}. According to the classification of EUV waves performed by \cite{warm11}, this EUV wave should be a nonlinear large-amplitude fast-mode wave that propagates faster than the ambient fast-mode speed and subsequently slow down due to decreasing amplitude. After the interaction of W2 with the bright points, the propagation speed of the EUV wave showed an obvious variation, of which the EUV wave accelerated (decelerated) significantly when passing through BP1 (BP2), and the propagation speed of the wave after interaction is about \speed{653} (\speed{354}) along C2 (C4). Since the propagation behavior of the EUV wave can not be distinguished on the imaging observations, we conjecture that the significant variation of the wave speed after the interaction with the bright points is caused by the changing of the propagating direction of the EUV wave during its interaction with the bright points, that is, due to the refraction or reflection effect. In addition, the speed (acceleration) of W2 along C4 at AIA 304 \AA\ is about \speed{637} (\acc{-159}), similar to the speed determined from AIA 193 \AA\ observations along the same path. The back-extrapolation of the fitted lines indicates that the initiation time of W2 should be  close to the onset time of the associated flare. This result seems to indicate a close relationship between the wave and the flare and thereby favors the scenario that EUV waves are driven by flares \citep[e.g.,][]{warm04b}. However, it is well known that the CME acceleration profile can be correlated with the flare lightcurve \citep[e.g.,][]{zhan01,temm08}. This is also consistent with the CME driven scenario of the EUV waves. In fact, the complex relationships among flares, CMEs and EUV waves are still unclear. In this sense, any apparent relation of the EUV wave onset time with the flare does not necessarily means that the flare is really related to the wave launch.

\begin{figure}\epsscale{0.8}
\plotone{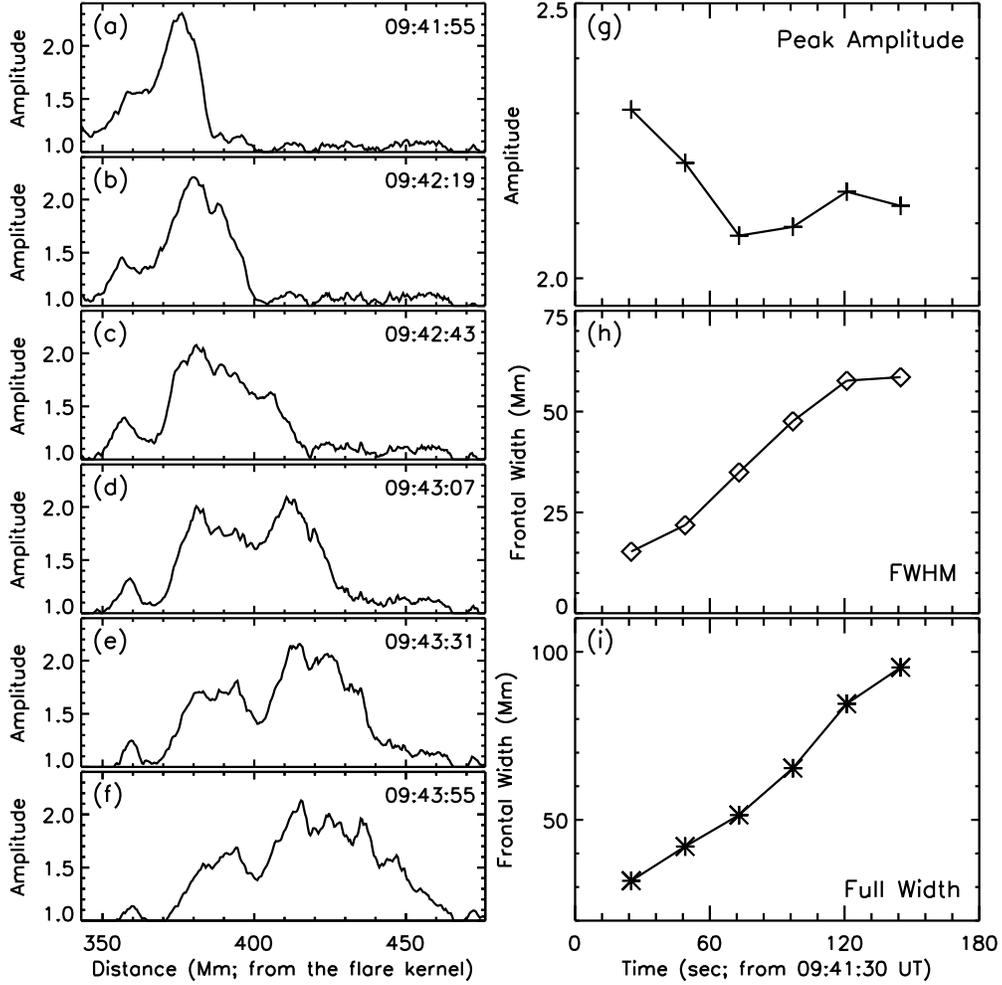}
\caption{Perturbation profiles (obtained from ratio images) of the most prominent part of the EUV wave ((a)--(f)), as indicated by the red ``$\times$'' symbols shown in \fig{fig4}(b). Note that all values that are smaller than one are set to one. Panel (g) shows the evolution of the maximum perturbation amplitude determined from panel (a)--(f). Panel (h) and (i) show the evolution of the full width at half-maximum and the full width of the frontal part of the perturbation profiles, respectively. \label{fig7}}
\end{figure}

The perturbation profiles of W2 along C3 are plotted in \fig{fig7}(a)--(f). These profiles are obtained from a time-distance diagram that is made from a series of AIA 193 \AA\ ratio images, where each frame is divided by a pre-event frame. It should be noted that we only show the profiles of the EUV wave within a distance from 350 to 480 Mm from the flare kernel, where the wave was most prominent as marked by the two red ``$\times$'' symbols in \fig{fig4}(b). The propagation of the EUV wave is well reflected by the evolution of these perturbation profiles. At 09:41:55 UT, the perturbation profile is the steepest and has the highest amplitude of about 2.3, corresponding to an enhancement of 130\% above the pre-event level. In the subsequent evolution, the amplitude of the wavefront decreases continuously with time. In the meantime, the width (steepness) of the wavefront shows significant broadening (declining). In principle, the AIA 193 \AA\ channel is an optically thin emission line, and the observed intensity should be proportional to the square of the plasma density and a function related to temperature. Therefore, the enhancement of intensity can result from the increase of both density and temperature. However, since EUV waves are usually observed in a wide temperature range simultaneously, one can assume that the intensity enhancement is primarily due to plasma compression rather than temperature changing. So $I/I_{\rm 0}=2.3$ in 193 \AA\ corresponds to a density ratio $n/n_{0} \varpropto (I/I_{\rm 0})^{1/2} \sim 1.5$. We also investigate the evolution of the amplitude, the frontal full width at half maximum, and the full width of the wavefront in panels (g), (h), and (i), respectively. The amplitude first decreases linearly and then keeps at a lower level, while the width increases monotonically during the entire evolution process. These results imply that the propagating EUV wave has a dispersive nature, confirming the results of \cite{vero10}, \cite{long11b}, and \cite{long11a}. It should be noted that these characteristic parameters of the wavefront might be affected by the changes in the quiet-Sun as well as by the brightenings induced by the wavefront passage such as the so-called ``stationary brightenings'' \citep{attr07}.

\section{CONCLUSION AND DISCUSSIONS}
By combining the high temporal and spatial resolution observations taken by {\em SDO} and {\em STEREO}, we investigate the detailed kinematics of the EUV wave that occurred on September 24, 2011. The EUV wave was accompanied by a {\em GOES} X1.9 flare and a fast partial halo CME. We find that there are several main arguments supporting the scenario that the observed large-scale propagating wave-like disturbances in the corona should be a compressive fast-mode magnetosonic wave, and it was driven by the CME expanding flanks during the initial stage of the EUV wave. Our main results are summarized as follows.

\begin{enumerate}
  \item The EUV wave propagated along a conical sector region of the quiet-Sun corona and can be observed in all EUV channels of the AIA across a wide temperature range. Similar kinematics of the EUV wave in these EUV channels are identified.
  \item The EUV wave (\speed{570}) propagated faster than the coronal sound speed in the quiet-Sun corona ($c_{\rm s}\leqslant$ \speed{200}), and the propagation showed a pronounced deceleration (\acc{-130}). This kinematical behavior of the EUV wave presented in this paper is consistent with the behaviors of the nonlinear large-amplitude waves that propagate faster than the ambient fast-mode magnetosonic speed and subsequently slow down due to decreasing amplitude \citep{warm11}.
  \item The wave natures of the EUV wave presented here are reflected through its interaction with other coronal magnetic structures along its propagation path, such as the oscillation of the loop and the filament, the reflection of the EUV wave from the remote polar coronal hole, and the variation of the propagation speed of the EUV wave due to the refraction or reflection effect of the bright points. These phenomena are not consistent with those pseudo-wave models such as by \cite{dela07,dela08} or \cite{attr07}. The fact that the wave keeps propagating after the dimming has stopped to expand actually rules out the pseudo-wave models for the current event.
  \item The dimming region following the wavefront stopped to expand at about 09:44 UT and the outer edge of the dimming region kept at a distance of about 50 Mm away from the flare kernel, but the EUV wave continued propagating out for a long distance until indiscernible on the visible disk. This result supports the scenario that the EUV waves are driven by the CME expanding flanks as suggested by many studies\citep{pats09a,muhr10,temm11}, in which the dimming region maps the CME footprint on the solar surface. After the driving of the CME expanding flanks, the EUV wave started its free propagation stage, during which the wavefront became more diffuse.
  \item Triangulation indicates that the propagation height of the EUV wave ranges from 60 to 100 Mm above the photosphere, which is consistent with \cite{pats09b} and \cite{kien09}.
  \item For the characteristics of the perturbation profile of the propagating wavefront, we find that both the steepness of the wavefront's leading edge and the amplitude of the wavefront decrease continuously with time. In the meantime, the width of the perturbation profile of the wavefront broadens during its evolution process. This broadening is observed for both the full width as well as for the full width at half-maximum. These observations indicate that the propagating EUV wavefront has a dispersive nature and could be described using a wave interpretation \citep{vero10,long11a}.
\end{enumerate}

Theoretically, the formation of the 304 \AA\ line is dominated by two chromospheric \ion{He}{2} lines at 303.781 and 303.786 \AA\, but a strong contribution from a coronal \ion{Si}{11} line at 303.32 \AA\ in the 304 \AA\ passband is also important, which has a peak formation temperature $\sim1.6$ MK \citep{bros96,pats09a,long11a}. Our observational results demonstrate that the kinematics of the EUV wave in 304 \AA\ observations has a similar behavior with other EUV lines, and the propagation height (60--100 Mm) of this EUV wave is higher than the height of the chromosphere. In addition, the wave signature is undetectable at the transition region line such as 1600 \AA. Thus we propose that the EUV wave presented in this paper is just a coronal phenomenon and no chromospheric counterpart. The wave signature seen in the 304 \AA\ passband is the manifestation of the contribution of the coronal \ion{Si}{11} line rather than the chromospheric \ion{He}{2} lines.

Secondary waves were also observed behind the primary fast wave front, which looks like the pseudo-wave predicted int the hybrid model of \cite{chen02}. However, in this hybrid model, the speed of the slower pseudo-wave should be about one third of the speed of the preceding fast shock wave. In the case presented in this paper, we find that the speed of the secondary waves was just slightly smaller than that of the primary fast wave. This discrepancy stimulates us to clarify the origin of these secondary waves. By checking the time-elapse movies made from the AIA data, we find that these secondary waves were probably generated by the interaction of the primary fast wave with the coronal structures on the wave propagation path, which is inconsistent with the slower pseudo-wave predict in the hybrid model presented by \cite{chen02}.

From the the profiles of the EUV wavefront, we obtain the highest amplitude $I/I_{\rm 0}=2.3$, which yields the peak density jump at the wavefront $n/n_{0} \sim 1.5$. This corresponds to a perpendicular fast magnetosonic wave Mach number of $M_{\rm ms}=1.4$ \citep{prie82}, which implies that the EUV wave presented in this paper should be a nonlinear fast-mode magnetosonic wave, confirming the results of \cite{vero10} and \cite{kien11}. Along with the evolution of the EUV wave, its amplitude decreases and thereby the Mach number of the EUV wave also decreases and approaches to a linear regime where the Mach number is close to unity 1. Here, the results including the deceleration to a Mach number of unity, amplitude decrease, and wave front broadening are all consistent with the nonlinear fast-mode wave scenario, as pointed out by \cite{warm04b}, and recently defined as one distinct class of coronal waves by \cite{warm11}.

In summary, our analysis results indicate that the EUV wave presented in this paper should be a nonlinear fast-mode magnetosonic wave that was driven by the CME expanding flanks during the initial stage, and then it propagated freely in the solar corona within a certain height ranging from 60 to 100 Mm above the photosphere. More investigations with high temporal and spatial resolution observations would be helpful to fully understand the physical nature of such propagating large-scale wave-like disturbances in the solar atmosphere.

\acknowledgments {\em SDO} is a mission for NASA's Living With a Star (LWS) Program. We thank the {\em STEREO}/SECCHI, {\em RHESSI}, and {\em GOES} teams for data support. We also thank the anonymous referee for constructive comments and suggestions that have improved the quality of this paper. This work is supported by the Natural Science Foundation of China under grants 10933003, 11078004, and 11073050, and the National Key Research Science Foundation (2011CB811400).

%\newpage

\begin{thebibliography}{33}
\expandafter\ifx\csname natexlab\endcsname\relax\def\natexlab#1{#1}\fi

\bibitem[Asai et al.(2012)]{asai12}
Asai, A., Ishii, T., Isobe, H., et al. 2012, \apjl, 745, L18
\bibitem[Aschwanden(2005)]{asch05}
Aschwanden, M. J. 2005, Physics of the Solar Corona: An Introduction with Problems and Solutions (2nd ed.; Chichester: Praxis Publishing Ltd.)
\bibitem[Attrill(2010)]{attr10}
Attrill, G. D. R. 2010, \apj, 718, 494
\bibitem[Attrill et al.(2009)]{attr09}
Attrill, G. D. R., Engell, A. J., Wills-Davey, M. J., Grigis, P., \& Testa, P. 2009, \apj, 704, 1296
\bibitem[Attrill et al.(2007)]{attr07}
Attrill, G. D. R., Harra, L. K., van Driel-Gesztelyi, L., \& D\'{e}moulin, P. 2007, \apj, 656, L101
%\bibitem[Balasubramaniam et al.(2010)]{bala10}
%Balasubramaniam, K. S., Cliver, E. W., Pevtsov, A., et al. 2010, \apj, 723, 601
\bibitem[Balasubramaniam et al.(2007)]{bala07}
Balasubramaniam, K. S., Pevtsov, A. A., \& Neidig, D. F. 2007, \apj, 658, 1372
\bibitem[Ballai et al.(2005)]{ball05}
Ballai, I., Erd\'{e}lyi, R., \& Pint\'{e}r, B. 2005, \apjl, 633, L145
\bibitem[Biesecker et al.(2002)]{bies02}
Biesecker, D. A., Myers, D. C., Thompson, B. J., Hammer, D. M., \& Vourlidas, A. 2002, \apj, 569, 1009
\bibitem[Brosius et al.(1996)]{bros96}
Brosius, J. W., Davila, J. M., Thomas, R. J., \& Monsignori-Fossi, B. C. 1996, \apjs, 106, 143
\bibitem[Brueckner et al.(1995)]{brue95}
Brueckner, G. E., Howard, R. A., Koomen, M. J., et al. 1995, \solphys, 162, 357
%\bibitem[Chen et al.(2010)]{chen10}
%Chen, F., Ding, M. D., \& Chen, P. F. 2010, \apj, 720, 1254
%\bibitem[Chen et al.(2011)]{chen11a}
%Chen, F., Ding, M. D., Chen, P. F., \& Harra, L. K. 2011, \apj, 740, 116
\bibitem[Chen(2006)]{chen06}
Chen, P. F. 2006, \apjl, 641, L153
%\bibitem[Chen(2009)]{chen09}
%Chen, P. F. 2009, \apjl, 689, L112
\bibitem[Chen et al.(2005)]{chen05}
Chen, P. F., Fang, C., \& Shibata, K. 2005, \apj, 622, 1202
\bibitem[Chen et al.(2002)]{chen02}
Chen, P. F., Wu, S. T., Shibata, K., \& Fang, C. 2002, \apjl, 572, L99
\bibitem[Chen \& Wu(2011)]{chen11}
Chen, P. F., \& Wu, Y. 2011, \apjl, 732, L20
\bibitem[Cheng et al.(2012)]{chen12}
Cheng, X., Zhang, J., Olmedo, O., et al. 2012, \apjl, 745, L5
\bibitem[Cliver et al.(2005)]{cliv05}
Cliver, E. W., Laurenza, M., Storini, M., \& Thompson, B. J. 2005, \apj, 631, 604
\bibitem[Cohen et al.(2009)]{cohe09}
Cohen, O., Attrill, G. D. R., Manchester IV, W. B., \& Wills-Davey, M. J. 2009, \apj, 705, 587
\bibitem[Dai et al.(2010)]{dai10}
Dai, Y., Auch\`{e}re, F., Vial, J.-C., Tang, Y. H., \& Zong, W. G. 2010, \apj, 708, 919
\bibitem[Delaboudini\`{e}re et al.(1995)]{dela95}
Delaboudini\`{e}re, J.-P., Artzner, G. E., Brunaud, J., et al. 1995, \solphys, 162, 291
\bibitem[Delann\'{e}e \& Aulanier(1999)]{dela99}
Delann\'{e}e, C., \& Aulanier, G. 1999, \solphys, 190, 107
\bibitem[Delann\'{e}e(2000)]{dela00}
Delann\'{e}e, C. 2000, \apj, 545, 512
\bibitem[Delann\'{e}e et al.(2007)]{dela07}
Delann\'{e}e, C., Hochedez, J.-F., \& Aulanier, G. 2007, \aap, 465, 603
\bibitem[Delann\'{e}e et al.(2008)]{dela08}
Delann\'{e}e, C., T\"{o}r\"{o}k, T., Aulanier, G., \& Hochedez, J.-F. 2008, \solphys, 247, 123
\bibitem[Domingo et al.(1995)]{domi95}
Domingo, V., Fleck, V., \& Poland, A. 1995, \solphys, 162, 1
\bibitem[Downs et al.(2011)]{down11}
Downs, C., Roussev, I. I., Van der Holst, B., et al. 2011, \apj, 728, 2
\bibitem[Eto et al.(2002)]{eto02}
Eto, S., Isobe, H., Narukage, N., et al. 2002, \pasj, 54, 481
\bibitem[Foley et al.(2003)]{fole03}
Foley, C. R., Harra, L. K., matthews, S. A., Culhane, J. L., \& Kitai, R. 2003, \aap, 399, 749
\bibitem[Gallagher \& Long(2011)]{gall11}
Gallagher, P. T., \& Long, D. M. 2011, \ssr, 158, 365
\bibitem[Gilbert et al.(2008)]{gilb08}
Gilbert, H. R., Daou, A. G., Young, D., Tripathi, D., \& Alexander, D. 2008, \apj, 685, 629
\bibitem[Gilbert \& Holzer(2004a)]{gilb04a}
Gilbert, H. R., \& Holzer, T. E. 2004a, \apj, 610, 572
\bibitem[Gilbert et al.(2004b)]{gilb04b}
Gilbert, H. R., Holzer, T. E., Thompson, B. J., \& Burkepile, J. T. 2004b, \apj, 607, 540
\bibitem[Gopalswamy et al.(2009)]{gopa09}
Gopalswamy, N., Yashiro, S., Temmer, M., et al. 2009, \apjl, 691, L123
\bibitem[Harra \& Sterling(2003)]{harr03}
Harra, L. K., \& Sterling, A. C. 2003, \apj, 587, 429
\bibitem[Hershaw et al.(2011)]{hers11}
Hershaw, J., Foullon, C., Nakariakov, V. M., \& Verwichte, E. 2011, \aap, 531, A53
\bibitem[Hudson et al.(2003)]{huds03}
Hudson, H. S., Khan, J. I., Lemen, J. R., Nitta, N. V., \& Uchida, Y. 2003, \solphys, 212, 121
\bibitem[Kaiser et al.(2008)]{kais08}
Kaiser, M. L., Kucera, T. A., Davila, J. M., et al. 2008, \ssr, 136, 5
\bibitem[Khan \& Aurass(2002)]{khan02}
Khan, J. I., \& Aurass, H. 2002, \aap, 383, 1018
\bibitem[Kienreich et al.(2009)]{kien09}
Kienreich, I. W., Temmer, M., \& Veronig, A. M. 2009, \apjl, 703, L118
\bibitem[Kienreich et al.(2011)]{kien11}
Kienreich, I. W., Veronig, A. M., Muhr, N., et al. 2011, \apjl, 727, L43
\bibitem[Klassen et al.(2000)]{klas00}
Klassen, A., Aurass, H., Mann, G., \& Thompson, B. J. 2000, A\&AS, 141, 357
\bibitem[Lemen et al.(2012)]{leme12}
Lemen, J. R., Title, A. M., Akin, D. J., et al. 2012, \solphys, 275, 17
\bibitem[Li et al.(2012)]{li12}
Li, T., Zhang, J., Yang, S., \& Liu, W. 2012, \apj, 746, 13
\bibitem[Lin et al.(2002)]{lin02}
Lin, R. P., Dennis, B. R., Hurford, G. J., et al. 2002, \solphys, 210, 3
\bibitem[Liu et al.(2010)]{liu10}
Liu, W., Nitta, N. V., Schrijver, C. J., Title, A. M., \& Tarbell, T. D. 2010, \apjl, 723, L53
%\bibitem[Liu et al.(2011)]{liu11}
%Liu, W., Title, A. M., Zhao, J., et al. 2011, \apjl, 736, L13
\bibitem[Long et al.(2011a)]{long11a}
Long, D. M., DeLuca, E. E., \& Gallagher, P. T. 2011a, \apjl, 741, L21
\bibitem[Long et al.(2008)]{long08}
Long, D. M., Gallagher, P. T., James McAteer, R. T., \& Bloomfield, S. 2008, \apjl, 680, L81
\bibitem[Long et al.(2011b)]{long11b}
Long, D. M., Gallagher, P. T., McAteer, R. T. J., \& Bloomfield, D. S. 2011b, \aap, 531, A42
\bibitem[Ma et al.(2009)]{ma09}
Ma, S., Wills-Davey, M. J., Lin, J., et al. 2009, \apj, 707, 503
\bibitem[Moreton(1960)]{more60}
Moreton, G. E. 1960, \aj, 65, 494
\bibitem[Moses et al.(1997)]{mose97}
Moses, D., Clette, F., Delaboudini\`{e}re, J.-P., et al. 1997, \solphys, 175, 571
\bibitem[Muhr et al.(2010)]{muhr10}
Muhr, N., Vr\v{s}nak, B., Temmer, M., Veronig, A. M., \& Magdaleni\'{c}, J. 2010, \apj, 708, 1639
\bibitem[Narukage et al.(2004)]{naru04}
Narukage, N., Morimoto, T., Kadota, M., et al. 2004, \pasj, 56, L5
\bibitem[Narukage et al.(2002)]{naru02}
Narukage, N., Hudson, H. S., Morimoto, T., et al. 2002, \apjl, 572, L109
\bibitem[Narukage et al.(2008)]{naru08}
Narukage, N., Ishii, T. T., Nagata, S., et al. 2008, \apjl, 684, L45
%\bibitem[Ofman et al.(2011)]{ofma11}
%Ofman, L., Liu, W., Title, A. M, \& Aschwanden, M. 2011, \apjl, 740, L33
\bibitem[Ofman \& Thompson(2002)]{ofma02}
Ofman, L., \& Thompson, B. J. 2002, \apj, 574, 440
\bibitem[Patsourakos \& Vourlidas(2009)]{pats09a}
Patsourakos, S., \& Vourlidas, A. 2009, \apjl, 700, L182
\bibitem[Patsourakos et al.(2009)]{pats09b}
Patsourakos, S., Vourlidas, A., Wang, Y. M., Stenborg, G., \& Thernisien, A. 2009b, \solphys, 259, 49
\bibitem[Podladchikova \& Berghmans(2005)]{podl05}
Podladchikova, O., \& Berghmans, D. 2005, \solphys, 228, 265
%\bibitem[Podladchikova et al.(2010)]{podl10}
%Podladchikova, O., Vourlidas, A., Van der Linden, A. M., W\"{u}lser, J.-P., \& Patsourakos, S. 2010, \apj, 709, 369
\bibitem[Priest(1982)]{prie82}
Priest, E. R. 1982, Solar Magneto-hydrodynamics (Dordrecht: Reidel)
\bibitem[Schmidt \& Ofman(2010)]{schm10}
Schmidt, J. M., \& Ofman, L. 2010, \apj, 713, 1008
\bibitem[Schou et al.(2012)]{scho12}
Schou, J., Borrero, J. M., Norton, A. A., et al. 2012, \solphys, 275, 327
\bibitem[Selwa et al.(2012)]{selw12}
Selwa, M., Poedts, S., \& DeVore, C. R. 2012, \apjl, 747, L21
\bibitem[Sheeley et al.(1999)]{shee99}
Sheeley, N. R., Walters, J. H., Wang, Y.-M., \& Howard, R. A. 1999, \jgr, 104, 24739
\bibitem[Temmer et al.(2011)]{temm11}
Temmer, M., Veronig, A. M., Gopalswamy, N., \& Yashiro, S. 2011, \solphys, 273, 421
\bibitem[Temmer et al.(2008)]{temm08}
Temmer, M., Veronig, A. M., Vr\v{s}nak, B., et al. 2008, \apjl, 673, L95
\bibitem[Thompson et al.(1999)]{thom99}
Thompson, B. J., Gurman, J. B., Neupert, W. M., et al. 1999, \apjl, 517, L151
\bibitem[Thompson \& Myers(2009)]{thom09}
Thompson, B. J., \& Myers, D. C. 2009, \apjs, 183, 225
\bibitem[Thompson et al.(1998)]{thom98}
Thompson, B. J., Plunkett, S. P., Gurman, J. B., et al. 1998, Geophys. Res. Lett. 25, 2465
\bibitem[Thompson et al.(2000)]{thom00}
Thompson, B. J., Reynolds, B., Aurass, H., et al. 2000, \solphys, 193, 161
\bibitem[Tripathi \& Raouafi(2007)]{trip07}
Tripathi, D., \& Raouafi, N.-E. 2007, \aap, 473, 951
\bibitem[Uchida(1968)]{uchi68}
Uchida, Y. 1968, \solphys, 4, 30
\bibitem[van Driel-Gesztelyi et al.(2008)]{van08}
van Driel-Gesztelyi, L., Attrill, G. D. R., D\'{e}moulin, P., Mandrini, C. H., Harra, L. K. 2008, ann. Geophys. 26, 3077
\bibitem[Veronig et al.(2011)]{vero11}
Veronig, A. M., G\"{o}m\"{o}ry, P., Kienreich, I. W., et al. 2011, \apjl, 743, L10
\bibitem[Veronig et al.(2010)]{vero10}
Veronig, A. M., Muhr, N., Kienreich, I. W., Temmer, M., \& Vr\v{s}nak, B. 2010, \apjl, 716, L57
\bibitem[Veronig et al.(2008)]{vero08}
Veronig, A. M., Temmer, M., \& Vr\v{s}nak, B. 2008, \apj, 681, L113
\bibitem[Veronig et al.(2006)]{vero06}
Veronig, A. M., Temmer, M., Vr\v{s}nak, B., \& Thalmann, J. K. 2006, \apj, 647, 1466
\bibitem[Vr\v{s}nak et al.(2001)]{vrsn01}
Vr\v{s}nak, B., Aurass, H., Magdaleni\'{c}, J., \& Gopalswamy, N. 2001, \aap, 377, 321
\bibitem[Vr\v{s}nak et al.(2005)]{vrsn05}
Vr\v{s}nak, B., Magdaleni\'{c}, J., Temmer, M., et al. 2005, 625, L67
\bibitem[Vr\v{s}nak et al.(2002)]{vrsn02}
Vr\v{s}nak, B., Warmuth, A., Braj\v{s}a, R., \& Hanslmeier, A. 2002, \aap, 394, 299
\bibitem[Vr\v{s}nak et al.(2006)]{vrsn06}
Vr\v{s}nak, B., Warmuth, A., Temmer, M., Veronig, A., Magdaleni\v{c}, J. 2006, \aap, 448, 739
\bibitem[Wang et al.(2009)]{wang09}
Wang, H., Shen, C., \& Lin, J. 2009, \apj, 700, 1716
\bibitem[Wang(2000)]{wang00}
Wang, Y.-M. 2000, \apjl, 543, L89
\bibitem[Warmuth(2010)]{warm10}
Warmuth, A. 2010, Adv. Space Res., 45, 527
\bibitem[Warmuth \& Mann(2011)]{warm11}
Warmuth, A., \& Mann, G. 2011, \aap, 532, A151
\bibitem[Warmuth \& Mann(2005)]{warm05a}
Warmuth, A., \& Mann, G. 2005a, \aap, 435, 1123
\bibitem[Warmuth et al.(2005)]{warm05b}
Warmuth, A., Mann, G., \& Aurass, H. 2005b, \apjl, 626, L121
\bibitem[Warmuth et al.(2001)]{warm01}
Warmuth, A., Vr\v{s}nak, B., Aurass, H., \& Hanslmeier, A. 2001, \apjl, 560, L105
\bibitem[Warmuth et al.(2004a)]{warm04a}
Warmuth, A., Vr\v{s}nak, B., Magdaleni\'{c}, J., Hanslmeier, A., \& Otruba, W. 2004a, \aap, 418, 1101
\bibitem[Warmuth et al.(2004b)]{warm04b}
Warmuth, A., Vr\v{s}nak, B., Magdaleni\'{c}, J., Hanslmeier, A., \& Otruba, W. 2004b, \aap, 418, 1117
\bibitem[West et al.(2011)]{west11}
West, M. J., Zhukov, A. N., Dolla, L., \& Rodriguez, L. 2011, \apj, 730, 122
\bibitem[White \& Thompson(2005)]{whit05}
White, S. M., \& Thompson, B. J. 2005, \apjl, 620, L63
\bibitem[Wills-Davey(2006)]{will06}
Wills-Davey, M. J. 2006, \apj, 645, 757
\bibitem[Wills-Davey \& Attrill(2009)]{will09}
Wills-Davey, M. J., \& Attrill, G. D. R. 2009, \ssr, 149, 325
\bibitem[Wills-Davey et al.(2007)]{will07}
Wills-Davey, M. J., DeForest, C. E., \& Stenflo, J. O. 2007, \apj, 664, 556
\bibitem[Wills-Davey \& Thompson(1999)]{will99}
Wills-Davey, M. J., \& Thompson, B. J. 1999, \solphys, 190, 467
\bibitem[Wu et al.(2001)]{wu01}
Wu, S. T., Zheng, H., Wang, S., et al. 2001, \jgr, 106, 25089
\bibitem[Wuelser et al.(2004)]{wuel04}
Wuelser, J., Lemen, J. R., Tarbell, T. D., et al. 2004, \procspie, 5171, 111
%\bibitem[Yang \& Chen(2010)]{yang10}
%Yang, H. Q., \& Chen, P. F. 2010, \solphys, 266, 59
%\bibitem[Yashiro et al.(2004)]{yash04}
%Yashiro, S., Gopalswamy, N., Michalek, G., et al. 2004, \jgr, 109, A07105
\bibitem[Zhang et al.(2001)]{zhan01}
Zhang, J., Dere, K. P., Howard, R. A., Kundu, M. R., \& White, S. M. 2001, \apj, 599, 452
\bibitem[Zheng et al.(2011)]{zhen11}
Zheng, R., Jiang, Y., Hong, J., et al. 2011, \apjl, 739L, 39
\bibitem[Zheng et al.(2012)]{zhen12}
Zheng, R., Jiang, Y., Yang, J., et al. 2012, \apj, 747, 67
\bibitem[Zhukov \& Auch\`{e}re(2004)]{zhuk04}
Zhukov, A. N., \& Auch\`{e}re, F. 2004, \aap, 427, 705
\bibitem[Zhukov et al.(2009)]{zhuk09}
Zhukov, A. N., Rodriguez, L., \& de Patoul, J. 2009, \solphys, 259, 73
\end{thebibliography}
\end{document}